\documentstyle{mn}

\title[CASE: SX Phe stars from the globular cluster $\omega$ Centauri]
{Cluster AgeS Experiment (CASE): ~SX Phe stars from
the globular cluster $\omega$ Centauri}
\author[A. Olech et. al.]
{A.~Olech$^{1}$, W.A. Dziembowski$^{2,1}$, A.A. Pamyatnykh$^{1,3}$,
J.~Kaluzny$^{1}$, W. Pych$^{1}$,
\newauthor
A. Schwarzenberg-Czerny$^{1,4}$ and I.B. Thompson$^{5}$\\
  $^1$Nicolaus Copernicus Astronomical Center,
     ul. Bartycka 18, 00-716 Warsaw, Poland
     (olech,wd,alosza,jka,pych,alex@camk.edu.pl)\\
  $^2$Warsaw University Observatory, Al. Ujazdowskie 4, 00-487
      Warszawa, Poland\\
  $^3$Institute of Astronomy, Russian Academy of Sciences,
     Pyatnitskaya Str. 48, 109017 Moscow, Russia\\
  $^4$Adam Mickiewicz University Observatory, ul. S{\l}oneczna 36,
60-286 Poznan, Poland\\
  $^5$Carnegie Institution of Washington, 813 Santa Barbara Street,
Pasadena,CA 91101, USA (ian@ociw.edu)
}
\begin{document}

\date{Accepted .................. Received ................ in original form ............}

\pagerange{\pageref{firstpage}--\pageref{lastpage}} \pubyear{2005}

\maketitle

\label{firstpage}

\begin{abstract}

We present an analysis and interpretation of oscillation spectra
for all 69 SX Phoenicis stars discovered in the field of the
cluster. For most of the stars we have reliable absolute magnitudes
and colors. Except of one, or perhaps two, objects, the stars are
cluster members. Their pulsational behaviour is very diversified.
Multiperiodic variability with at least part of the excited modes
being nonradial is most common but there are also many cases of
high amplitude, presumably radial mode, pulsators. In a number of
such cases we have evidence for two radial modes being excited.
Parameters of radial mode pulsators are in most cases consistent
with standard evolutionary models for stars in the mass range
$0.9\div1.15M_\odot$. However, in four cases we have evidence
that the masses are significantly lower than expected. Three
objects show frequency triplets that may be interpreted in terms
of rotational frequency splitting of $\ell=1$ modes. Implied
equatorial velocities of rotation are from 10 to over 100 km/s.
Nearly all measured frequencies fall in the ranges predicted for
unstable modes. Two cases of low frequency variability are
interpreted as being caused by tidal distortion induced by close
companions.

\end{abstract}

\begin{keywords}
stars: SX Phe - stars: variables -- globular clusters: individual:
$\omega$ Cen
\end{keywords}

\section{Introduction}

SX Phoenicis stars (SXPS) are the Population II short period
pulsators, in most respects similar to much more numerous $\delta$
Scuti stars, which are the Population I objects. Both types occupy
the low luminosity end of the Cepheid instability strip in the H-R
diagram. It is not clear whether there are any systematic
differences in pulsation properties between the two types. The
incidence of high amplitude pulsation once seemed much higher
among SXPS but now, with improved cluster photometry, many low
amplitude SXPS are being detected (Pych et al. 2001, Mazur et al.
2003, Kaluzny \& Thompson 2003, Kaluzny et al. 2004).

What makes SXPS interesting in a wider context is that they are
blue straggler stars (BSS) found in a large number in globular and
old open clusters. This means that they have unusual life history
which is not well understood as it cannot be explained in terms of
the standard single star evolution scenario. The problem of BSS
origin is being debated for decades. Historically, the first
explanation was that the objects are products of evolution of the
mass receiving component in close binary systems (McCrea 1964).
Later scenarios involving mergers of main sequence stars were
mostly considered. The scenarios included mergers of primordial
binaries after a gradual decrease of the orbit (e.g. Carney et al.
2001) as well as direct stellar collisions in dense cluster cores
(e.g. Lombardi et al. 2002). It is not clear how much imprints of
the past evolution should be left in individual BSS. This depends
on the efficiency of chemical element and angular momentum mixing
- the processes still poorly understood in stellar interior
physics. Understanding the origin and internal structure of BSS is
a challenging task for stellar evolution theory. The answer is of
interest not only for this field but also for globular cluster
research. As emphasized by Lombardi \& Rasio (2002), collisions
and mergers of stars which lead to BSS formation, play an
important role in evolution of these systems.

Pulsation data on SXPS are potential sources of accurate
constraints on models of BSS. Each measured frequency of an
identified oscillation mode is such a constraint. Admittedly, mode
identification is difficult unless we have evidence that the
excited modes are radial. Double mode radial pulsators seem quite
frequent among SXPS.  The two accurately measured numbers yield
strong constraints on stellar mass and heavy element content in
the interior. Data on nonradial mode frequencies could be even
more interesting. In particular, determination of the frequency
splitting  provides certain mean value of the rotation rate in the
interior. Furthermore, there are nonradial modes whose frequencies
are very sensitive to the extent of the element mixing beyond the
convective core boundary. Of great interest in the context of the
debated BSS origin is seeking evidence for presence of close
companions manifesting themselves through cyclic period variations
or tidally induced variability.

What makes SXPS interesting for stellar pulsation theory is the
large diversity of their pulsation forms. In a relatively narrow
period range, extending from less than half to two hours, we
encounter both the multimode low-amplitude pulsation, typical for
main sequence dwarfs, and the monomode high-amplitude pulsation,
typical for Cepheids. According to linear nonadiabatic
calculations, there are many unstable modes in SXPS. However,
finite amplitude development of the instability is not understood.
This is the most outstanding unsolved problem of the theory. A
sample of SXPS with well constrained mean parameters may provide a
key information what makes a star to become a dwarf- or a
giant-type pulsator.

The globular cluster $\omega$ Centauri houses the largest number
of SXPS of all systems in our Galaxy (Kaluzny et al. 2004). Having
a large number of objects with well determined luminosities is an
obvious advantage. The fact that the stars cannot be assumed
coeval and of the same chemical composition (Rey et al. 2000) is a
complicating factor. The cluster is atypical. Data on SXPS may
prove useful in disentangling its evolution.

In the next section we survey observational data on all SXPS in $\omega$
Cen which includes frequency analysis. In section 3 we compare mean
photometric parameters of SXPS with the corresponding values calculated
for standard evolutionary models. We also provide information on
pulsation properties of these models based on the linear nonadiabatic
analysis. Section 4 is a star by star analysis of individual objects in
which we compare their observational properties with theoretical models.
In section 5 we summarize our results.

\section{Photometry}\label{s2}

$\omega$ Centauri contains the most numerous population of
variable stars among globular clusters of the Galaxy (Clement et
al. 2001). However until the mid 1990ies there was only one SX
Phe-type variable known in the field of the cluster. It was V65,
which in fact turned out later to be a foreground star. However,
not much later Kaluzny et al. (1996, 1997) reported a discovery of
25 SXPS, most of which must be members of the cluster. Subsequent
work done by Kaluzny et al. (2004) increased the number of SXPS in
the field of $\omega$ Centauri to 69, making it the richest in
these variables among all globular clusters of our Galaxy.

Kaluzny et al. (2004) provide extensive photometry for 61 SXPS collected
during two years. This photometry contains from 532 to 755 $V$ and over
150 $B$ measurements for each variable. It is an excellent data base for
analyzing multi-mode behavior and extracting information on physical
properties of individual objects.

\begin{table}
 \centering
 \begin{minipage}{200mm}
  \caption{Basic properties of SX Phe-type variables in field
           of $\omega$ Cen.\label{t1}}
{\tiny
  \begin{tabular}{llccccc}
\hline
Star & Period & Amp. & $<V>$ & $B-V$ & $M_V$ & $(B-V)_0$\\
\hline
V65 & 0.0627235 & 0.18 & 14.922 & 0.361 & 0.832 & 0.231\\
V194 & 0.0471777 & 0.51 & 17.016 & 0.330 & 2.926 & 0.200\\
V195 & 0.0654912 & 0.38 & 16.780 & 0.371 & 2.690 & 0.241\\
V196 & 0.05740 & 0.23 & 17.000 & - & 2.910 & - \\
V197 & 0.0471210 & 0.13 & 16.850 & 0.546 & 2.760 & 0.416\\
V198 & 0.0481817 & 0.15 & 17.533 & 0.323 & 3.483 & 0.193\\
V199 & 0.0622867 & 0.73 & 16.689 & 0.333 & 2.599 & 0.203\\
V200 & 0.0495210 & 0.28 & 16.568 & 0.541 & 2.478 & 0.411\\
V201 & 0.05065 & 0.19 & 17.200 & - & 3.110 & - \\
V202 & 0.04642 & 0.12 & 17.170 & - & 3.080 & - \\
V203 & 0.04178 & 0.25 & 16.750 & - & 2.660 & - \\
V204 & 0.0493757 & 0.40 & 16.881 & 0.363 & 2.791 & 0.233\\
V217 & 0.0532609 & 0.10 & 17.038 & 0.429 & 2.948 & 0.299\\
V218 & 0.0437393 & 0.07 & 17.095 & 0.330 & 3.005 & 0.200\\
V219 & 0.0386680 & 0.08 & 17.303 & 0.337 & 3.213 & 0.207\\
V220 & 0.0528868 & 0.12 & 16.986 & 0.357 & 2.896 & 0.227\\
V221 & 0.0361336 & 0.05 & 16.680 & 0.451 & 2.590 & 0.321\\
V222 & 0.03891 & 0.05 & 17.310 & - & 3.220 & - \\
V225 & 0.0486381 & 0.22 & 16.845 & 0.393 & 2.755 & 0.262\\
V226 & 0.0378523 & 0.17 & 17.299 & 0.413 & 3.209 & 0.283\\
V227 & 0.0382255 & 0.05 & 17.272 & 0.397 & 3.182 & 0.267\\
V228 & 0.0398531 & 0.08 & 17.199 & 0.403 & 3.109 & 0.273\\
V229 & 0.0375333 & 0.09 & 17.407 & 0.294 & 3.317 & 0.164\\
V230 & 0.03388 & 0.03 & 16.550 & - & 2.460 & - \\
V231 & 0.0374845 & 0.05 & 17.419 & 0.341 & 3.329 & 0.211\\
V232 & 0.03697 & 0.04 & 17.590 & - & 3.500 & - \\
V233 & 0.0365376 & 0.10 & 17.210 & 0.353 & 3.120 & 0.223\\
V237 & 0.0656024 & 0.27 & 16.861 & 0.351 & 2.771 & 0.221\\
V238 & 0.0408004 & 0.07 & 17.355 & 0.349 & 3.265 & 0.219\\
V249 & 0.0349468 & 0.10 & 17.435 & 0.361 & 3.345 & 0.231\\
V250 & 0.0406269 & 0.07 & 17.433 & 0.380 & 3.343 & 0.250\\
V252 & 0.0466226 & 0.06 & 17.445 & 0.405 & 3.355 & 0.275\\
V253 & 0.0399687 & 0.11 & 17.232 & 0.308 & 3.142 & 0.178\\
V260 & 0.04626 & 0.05 & 17.080 & - & 2.990 & - \\
NV294 & 0.01773360 & 0.02 & 17.292 & 0.328 & 3.202 & 0.198\\
NV295 & 0.01823127 & 0.01 & 17.283 & 0.357 & 3.193 & 0.227\\
NV296 & 0.02212644 & 0.03 & 16.946 & 0.365 & 2.856 & 0.235\\
NV297 & 0.03389566 & 0.02 & 16.628 & 0.301 & 2.538 & 0.171\\
NV298 & 0.0330389 & 0.08 & 17.410 & 0.354 & 3.320 & 0.224\\
NV299 & 0.0344409 & 0.04 & 17.325 & 0.432 & 3.235 & 0.302\\
NV300 & 0.0347301 & 0.02 & 17.484 & 0.363 & 3.394 & 0.233\\
NV301 & 0.0354431 & 0.03 & 16.974 & 0.439 & 2.884 & 0.309\\
NV302 & 0.0355192 & 0.04 & 17.081 & 0.362 & 2.991 & 0.232\\
NV303 & 0.0359503 & 0.01 & 16.932 & 0.322 & 2.842 & 0.192\\
NV304 & 0.0361405 & 0.03 & 17.236 & 0.421 & 3.146 & 0.291\\
NV305 & 0.0365672 & 0.04 & 17.384 & 0.391 & 3.294 & 0.261\\
NV306 & 0.0384044 & 0.06 & 17.528 & 0.370 & 3.438 & 0.240\\
NV307 & 0.0385032 & 0.07 & 17.069 & 0.519 & 2.979 & 0.389\\
NV308 & 0.0389852 & 0.05 & 17.278 & 0.382 & 3.188 & 0.252\\
NV309 & 0.0397455 & 0.04 & 16.591 & 0.362 & 2.501 & 0.232\\
NV310 & 0.0401776 & 0.02 & 16.791 & 0.415 & 2.701 & 0.285\\
NV311 & 0.0414132 & - & - & - & - & - \\
NV312 & 0.0433272 & 0.06 & 16.394 & 0.390 & 2.304 & 0.26\\
NV313 & 0.0418484 & 0.16 & 17.678 & 0.348 & 3.588 & 0.218\\
NV314 & 0.0421220 & 0.08 & 17.086 & 0.386 & 2.996 & 0.256\\
NV315 & 0.0422811 & 0.10 & 16.392 & 0.518 & 2.302 & 0.388\\
NV316 & 0.0424040 & 0.03 & 17.326 & 0.326 & 3.236 & 0.196\\
NV317 & 0.0426396 & 0.05 & 16.968 & 0.453 & 2.878 & 0.323\\
NV318 & 0.0437306 & 0.02 & 16.799 & 0.472 & 2.709 & 0.342\\
NV319 & 0.0489421 & 0.10 & 17.239 & 0.465 & 3.149 & 0.335\\
NV320 & 0.0471936 & 0.08 & 17.294 & 0.440 & 3.204 & 0.310\\
NV321 & 0.0474854 & 0.10 & 16.409 & 0.136 & 2.319 & 0.006\\
NV322 & 0.0479562 & 0.08 & 17.096 & - & 3.006 & - \\
NV323 & 0.0493547 & 0.03 & 16.638 & 0.436 & 2.548 & 0.307\\
NV324 & 0.0512944 & 0.24 & 16.402 & 0.297 & 2.312 & 0.167\\
NV325 & 0.0535445 & 0.10 & 16.410 & 0.501 & 2.320 & 0.371\\
NV326 & 0.0569058 & 0.18 & 17.041 & 0.401 & 2.951 & 0.271\\
NV327 & 0.0606414 & 0.09 & 16.642 & 0.274 & 2.552 & 0.144\\
NV328 & 0.0899030 & 0.02 & 17.054 & 0.598 & 2.964 & 0.468\\
\hline
\end{tabular}}
\end{minipage}
\end{table}

In addition, $\omega$ Cen is the first globular cluster for which the
CASE project (Thompson et al. 2001) determined the distance. The
analysis of photometric and spectroscopic data for an eclipsing binary
OGLE GC17 yielded an apparent distance modulus of $(m-M)_V=14.09\pm0.04$
mag (Kaluzny et al. 2002). It is the most precise and reliable distance
determination for this cluster and will be used in our work for
calculating the absolute magnitudes of SXPS belonging to $\omega$ Cen.
We will also use the mean color data dereddened with the color excess
$E(B-V)=0.13$ mag (Schlegel, Finkbeiner \&  Davis, 1998) to place
individual objects in the H-R diagram.

The full list of SXPS located in the field of $\omega$ Centauri
containing main periods, amplitudes, mean magnitudes and colors is
provided in Table 1. Fig. 1 shows color-magnitude diagram around
blue stragglers region of $\omega$ Cen. Points and open circles
denote constant stars and SXPS, respectively. Not all stars in the
SXPS region are variables. This is hardly surprising, as we have
the same situation in the $\delta$ Scuti domain. Presumably these
apparently constant objects are very low amplitude pulsators.

\begin{figure}
\begin{center}
\vspace{9.9cm}
\end{center}
\caption{Color-magnitude diagram around blue stragglers region of
$\omega$ Cen. Points and open circles denote constant stars and
SXPS, respectively.}
\includegraphics{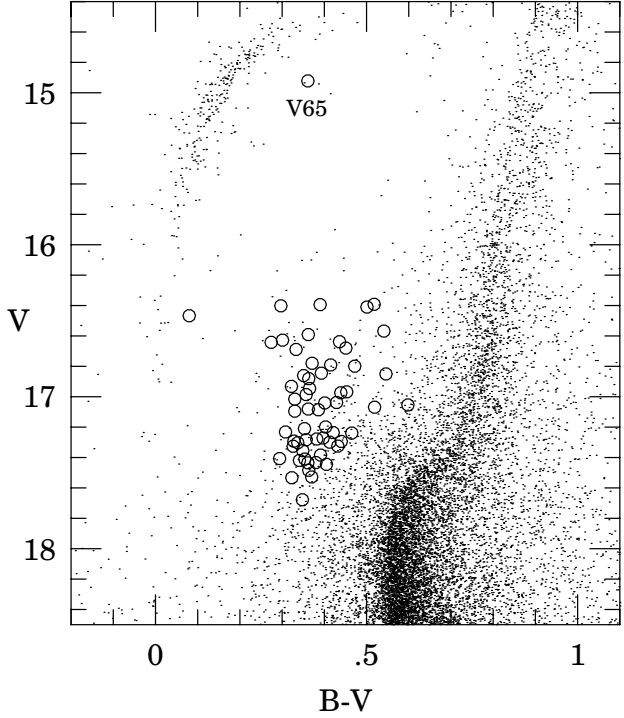}
\end{figure}

\begin{figure}
\vspace{21cm} \caption{The uppermost panel shows the ANOVA
periodogram of the raw light curve of NV324. The panels below show
spectra after consecutive prewhitenings.} \includegraphics{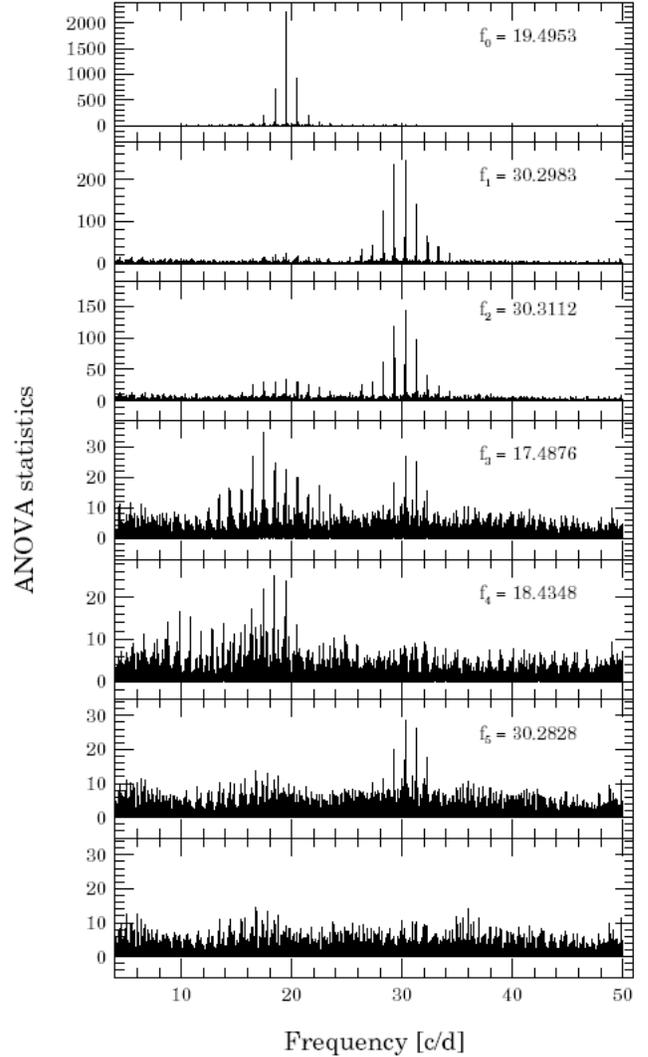}
\end{figure}

\begin{figure}
\vspace{14cm}
\caption{The light curves of NV324 associated with all six periods found.
Each light curve, phased with the particular period, was prewhitened with
six remaining periods and their harmonics.}
\includegraphics{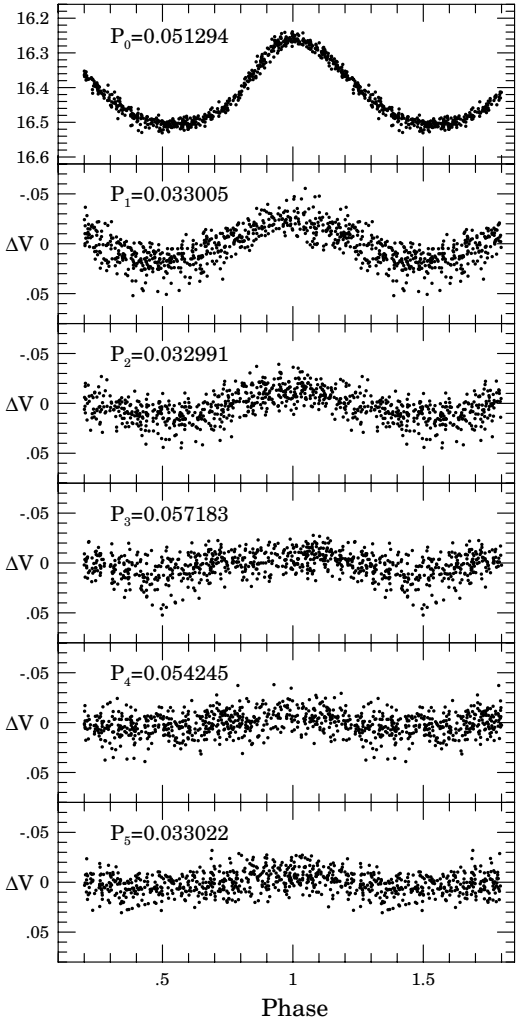}
\end{figure}

\section{Frequency analysis}\label{s3}

Many SX Phe stars exhibit multiple periods hence their analysis
poses specific challenges. Arguably the most advanced project in
terms of both extent of its photometry and multitude of detected
periods, the Whole Earth Telescope (WET), employed either the
prewhitening {\em or} synthesis methods in their analysis (Kepler,
1993). From the statistical point of view the two methods
correspond to Gauss-Seidel (G-S) and Newton-Raphson (N-R)
solutions of the least squares problem (LSQ). On one hand, a
current astronomical practice favors the synthesis (N-R) method,
employing the covariance matrix with large extra-diagonal
coefficients. Presence of close frequencies and/or their
combinations in the synthesis method yields near singular normal
equations and may produce large near cancelling terms in the
solution, creating the $\infty-\infty$ problem and yielding
solutions with excess amplitudes. On the other hand, the method
recommended to deal with singularity in least squares fitting
(LSQ) is by means of the singular value decomposition (SVD, e.g.
Press et al., 1986). In the SVD procedure the near singular terms
of the covariance matrix are set small, resulting in choice of the
solution with minimum norm (minimum amplitudes in the present
context). This resembles somewhat the prewhitening method, where
implicitly all extra-diagonal terms are set zero (Gray \&
Desikachary, 1973). We are unaware of any paper comparing merits
of the prewhitening and synthesis methods in rigid mathematical
(statistical) terms.

We employ the prewhitening and synthesis methods in parallel to
analyze our SX Phe light curves. We employed no more than 6
harmonics in prewhitening and in most cases just 2 or 3 harmonics
for the $f_0$ mode and one harmonic for other modes.  The values of
(fundamental) frequencies were adjusted by nonlinear iterations,
their harmonics/combinations were tied to the base frequencies.
The resulting frequencies are listed in Table 2. The both
methods yielded consistent frequencies for all strong detections,
corresponding to the AoV statistics $\Theta>15$. We observed no
adverse effect of reappearance of the frequencies once
prewhitened. Ambiguity, if any,  arose occasionally due to
aliasing and it manifested with equal strength in both the
prewhitening and synthesis methods. Such cases are marked in Table
2 with superscript $^1$.

The novelty in our approach, as compared with WET, is in use of the
multi-harmonic AoV periodogram (Schwarzenberg-Czerny, 1996). It employs
orthogonal functions and is able to combine power from harmonics (NH=2
harmonics were combined in practice). Importance of use of orthogonal
functions in period search was argued by Lomb (1976), Ferraz-Mello
(1981), Scargle (1982) and Foster (1994). Advantage in combining
harmonics is two-fold: they contribute extra power for the base
frequency and at the same time reduce power in the residuals. Any method
sensitive to the harmonics (e.g. phase folding and binning and the
present case of Fourier fit) is prone to produce sub-harmonics in the
periodogram. The subharmonic pose little problem in practice as they are
easily identified by tight packing of aliases. There exist exact
relations between power in the data, root-mean square (rms) of the
residuals and the AoV statistics $\Theta$ employed in the present case
(e.g. Schwarzenberg-Czerny, 1999). Detection of frequencies poses
special case of hypothesis testing in statistics. In our case the
detection criterion $\Theta>\Theta_c\equiv 8$ roughly corresponded to
mode amplitude exceeding 4 times its standard deviation $A>4\sigma_A$.
Any frequencies near/below that detection limit are deemed unsafe and
indicated with superscript $^2$ in Table 2. The purpose of listing these
marginal detections ($15>\Theta>8$) is solely to indicate deviations of
residuals from pure noise and not to ponder on frequencies.

An example of our results for NV324 is shown in Fig. 2. The panels show
ANOVA periodograms (Schwarzenberg-Czerny 1996). The uppermost panel is
computed for the original data and reveals the dominant peak with its
aliases. The remaining ones are computed for the data after subtraction
of all so far identified frequencies and significant combinations of
thereof. Fig. 3 shows the light curves of NV324 associated with all six
periods found. Each light curve, phased with the particular period, was
prewhitened with five remaining periods and their harmonics. Note the
asymmetry of the three upper light curves which is typical for pulsating
stars. The window function is a concept associated with the power
spectrum. Strictly speaking it does not apply for the ANOVA periodogram.
However, in Fig. 2 numerous window-like structures are present around
principal frequencies.

We refrain here from discussion of amplitudes. On one hand theory
results on amplitudes are unreliable (Sect. \ref{s4}). On the
other hand LSQ fits of multifrequency light curves often suffer
from large correlation of parameters, indicating large error
ellipses despite small variances. We address this problem only
indirectly here by relying our analysis on a large sample of stars
with very extensive coverage of observations. Application of two
independent reduction methods benefitted us in that we checked
against any computation errors as well as gained insight into
repeatability and reliability of our results.

\begin{table*}
 \centering
 \begin{minipage}{200mm}
  \caption{The frequencies found in the light curves of SX Phe variables
          in the field of $\omega$ Cen.\label{t2}}
  \begin{tabular}{lccccccc}
\hline
Star & $f_0$ & $f_1$ & $f_2$ & $f_3$ & $f_4$ & $f_5$ & $f_6$ \\
\hline
V65 & 15.9430(2) & 20.2287(1) & 21.2182(8) & 24.5199(8) &
36.1720(8)$^1$ & - & - \\ V194 & 21.1964(1) & 27.1633(5) &
43.7932(9) & 20.7471(6)$^1$ & 22.1462(6)$^2$ & - & - \\ V195 &
15.2692(1) & 15.2747(5) & 15.2668(6) & - & - & - & - \\ V197 &
21.2219(2) & 21.3904(5) & 21.3129(5) & 27.1838(8) & - & - & - \\
V198 & 20.7548(3) & - & - & - & - & - & - \\ V199 & 16.0548(2) & -
& - & - & - & - & - \\ V200 & 20.1935(1) & 26.5308(5) & - & - & -
& - & - \\ V204 & 20.2529(1) & 26.1552(5) & - & - & - & - & - \\
V217 & 18.7755(5) & - & - & - & - & - & - \\ V218 & 22.8627(5) &
27.0008(6)$^1$ & 43.6865(9) & - & - & - & - \\ V219 & 25.8611(6) &
26.6204(8) & 28.2532(8) & 3.2836(3) & - & - & - \\ V220 &
18.9083(2) & 24.3161(5) & 23.8450(7) & - & - & - & - \\ V221 &
27.6749(4) & 26.9153(5) & 32.8848(5) & 27.6867(6) & 27.6630(9) & -
& - \\ V225 & 20.5600(2) & 26.4145(5) & 26.8212(6) & - & - & - & -
\\ V226 & 26.4185(5) & 27.4932(6) & 15.4167(5) & 25.4523(8) & - &
- & - \\ V227 & 26.1605(5) & 25.4535(9) & 28.6334(8) & - & - & - &
- \\ V228 & 25.0923(5) & 24.3732(7) & 26.7906(7)$^1$ & 30.0974(9) & -
& - & - \\ V229 & 26.6429(3) & 51.6073(6) & - & - & - & - & - \\
V231 & 26.6778(4) & 27.1269(6) & 27.3509(7) & 36.6960(6)$^2$ & - &
- & - \\ V233 & 27.3691(2) & 2.9205(5) & - & - & - & - & - \\ V237
& 15.2433(2) & 24.1162(5) & 15.2013(8) & - & - & - & - \\ V238 &
24.5096(4) & 24.1318(8) & 25.9960(7) & 24.2163(7)$^1$ &
39.5205(7)$^2$ & - & - \\ V249 & 28.6149(4) & 35.0000(9) &
29.6107(8) & 28.6212(8) & - & - & - \\ V250 & 24.6142(3) &
37.6484(8) & - & - & - & - & - \\ V252 & 21.4488(4) & - & - & - &
- & - & - \\ V253 & 25.0196(5) & 25.7372(6) & 33.8173(6) &
31.1610(7)$^1$ & - & - & - \\ NV294 & 56.3901(6) & 28.9864(6) &
52.0186(6) & 48.4195(6)$^{1,2}$ & - & - & - \\ NV295 & 54.8509(5)
& 28.3558(6) & 53.9590(6)$^1$ & 31.5356(6)$^1$ & - & - & - \\
NV296 & 45.1948(7) & 40.3318(7)$^1$ & 38.6563(7) & 42.3111(8) &
30.5764(7)$^2$ & - & - \\ NV297 & 29.5022(7) & 30.7973(7) &
31.8181(7) & 30.6969(9) & - & - & - \\ NV298 & 30.2674(4) &
31.5946(4) & - & - & - & - & - \\ NV299 & 29.0352(6) & 33.2050(8)
& - & - & - & - & - \\ NV300 & 29.7935(6) & - & - & - & - & - & -
\\ NV301 & 28.2141(5) & 27.8536(6) & 35.3744(7) & 51.6717(9) & - &
- & - \\ NV302 & 28.1538(7) & - & - & - & - & - & - \\ NV303 &
27.8162(7) & - & - & - & - & - & - \\ NV304 & 27.6698(8) & - & - &
- & - & - & - \\ NV305 & 27.3469(5) & 22.6529(6) & 19.0827(7) &
15.9932(7)$^2$ & - & - & - \\ NV306 & 26.0387(5) & 40.7315(6) & - & -
& - & - & - \\ NV307 & 25.9719(5) & - & - & - & - & - & - \\ NV308
& 25.6507(5) & 26.3387(5) & 26.0989(7)$^1$ & 49.6777(8) &
42.2776(8)$^2$ & - & - \\ NV309 & 25.1601(6) & 37.8653(9) &
36.0886(8) & 25.5930(8)$^1$ & 30.2689(9)$^1$ & 24.5741(9)$^2$ &
38.8636(7)$^1$\\ NV310 & 24.8894(8) & 25.9192(6) & - & - & - & - &
- \\ NV311 & 24.1469(6) & 36.6940(5) & - & - & - & - & - \\ NV312
& 23.0802(5) & 36.7987(8) & - & - & - & - & - \\ NV313 &
23.8958(4) & - & - & - & - & - & - \\ NV314 & 23.7405(5) &
23.7468(5) & 36.4025(6) & - & - & - & - \\ NV315 & 23.6513(5) &
23.1521(4) & 22.8341(6) & 34.3549(6)$^1$ & - & - & \\ NV316 &
23.5827(8) & - & - & - & - & - & - \\ NV317 & 23.4525(5) &
22.7566(6)$^1$ & 24.9835(6) & 24.4356(7) & - & - & - \\ NV318 &
22.8673(7) & - & - & - & - & - & - \\ NV319 & 20.4323(4) & - & - &
- & - & - & - \\ NV320 & 21.1894(5) & - & - & - & - & - & - \\
NV321 & 21.0591(5) & 21.5337(7) & 20.1156(9) & - & - & - & - \\
NV322 & 20.8524(6) & 26.8419(6) & 20.4605(6) & 27.5321(7)$^1$ &
43.3720(6)$^1$ & 25.0637(9)$^1$ & 35.9001(6)$^1$\\ NV323 &
20.2615(6) & 15.7478(6) & - & - & - & - & - \\ NV324 & 19.4953(4)
& 30.2983(7) & 30.3112(9) & 17.4876(8) & 18.4348(9)$^1$ &
30.2828(7)$^1$ & - \\ NV325 & 18.6761(5) & 19.0173(7) & 19.0683(5)
& 36.4063(7) & 36.3059(6)$^1$ & 34.6931(7)$^1$ & - \\ NV326 &
17.5728(6) & 22.5375(6) & - & - & - & - & - \\ NV327 & 16.4904(3)
& 33.2019(9) & - & - & - & - & - \\ NV328 & 11.1231(6) & - & - & -
& - & - & - \\ \hline \multicolumn{8}{l}{$^1$ - problems with
aliasing,  ~$^2$ - close to the detection limit}
\end{tabular}
\end{minipage}
\end{table*}

\section{Evolutionary models and their oscillation properties}\label{s4}

Model calculations were made with Warsaw-New Jersey stellar
evolution code which is a modern version of Paczy\'nski's (1970)
code developed mainly by M. Koz{\l}owski and R. Sienkiewicz to
include new opacity and EOS data -both from the OPAL project
(Iglesias \& Rogers 1996, Rogers, Swenson \& Iglesias 1996,
respectively) - mean effect of rotation, and overshooting.
Oscillation properties were calculated with modernized version of
Dziembowski's (1977) code, which includes effects of rotation.

Our default hypothesis is that stellar structure is adequately
described by standard stellar models, that is that there is no
imprints of mass exchange or merging in their current structure.
When discussing individual objects we ask whether the data are
consistent with this hypothesis. The test is possible if we have a
plausible identification of at least one peak in the oscillation
spectra with a radial mode. If a measured period is found longer
than that of fundamental mode in the model consistent with the
star position in the H-R diagram and this position also excludes
driving of $g$-modes, then we may suspect that the object has
lower mass than implied by its luminosity.

We consider interpretation of three close frequencies in terms of
rotational splitting of $\ell=1$ mode frequencies. The
self-consistency of the hypothesis is checked by comparing the
observed and calculated multiplet structures. The frequency
distance between the extreme components is a measure of mean
rotation rate, which is mode-dependent because different modes
probe different parts of the star. The consistency test is
possible if nearly uniform rotation is assumed. Then we can
evaluate the departure from symmetry induced by higher order
effects of rotation and compare the result with observations.

Even if we cannot identify individual peaks in oscillation
spectra, still we can compare the frequency ranges where the peaks
occur with the range of unstable modes in the models. This is of
interest because the presence of oscillations left of the blue
edge may indicate a substantial helium enhancement, which is
expected in certain models of BSS formation.

Peaks at frequencies much lower than that of the fundamental radial mode
are of our interest too. These could be attributed to high order
$g$-modes. It is now a debated problem whether such modes may be excited
in $\delta$ Scuti stars (Handler 1999, Breger et al. 2002, Dupret et al. 2004).
Even more interesting possibility is
that such peaks represent tidally induced changes (Aerts et al. 2002, Handler
et al. 2002).

Figure 4 shows evolutionary tracks and positions of SXPS from $\omega$
Cen in the color-magnitude diagram. Tracks were calculated with two sets
of the chemical composition  parameters $(Z=0.002, X=0.74)$ and
$(Z=0.0002, X=0.75)$. The range of $Z$ values is consistent with metal
abundances observed among the stars belonging to the cluster (Rey et el.
2000). The range of masses, which is given in the caption, was chosen to
cover the range of stellar $B - V$ and $M_V$ values. The tracks were
converted from the $\log T_{\rm eff}$ -- $\log L$  to the $(B-V)$ --
$M_V$ diagram with the use of Kurucz's (1998) tabular data based on his
stellar atmosphere model calculations.

\begin{figure}
\vspace{9cm}
\caption{The H-R diagram showing evolutionary tracks in the mass
range $1.0\div1.2$ with $0.05 M_\odot$  step at $Z=0.002$ (dashed line) and
$0.85\div1.05$ at $Z=0.0002$ (solid line).
Straight lines show corresponding positions of the first overtone blue edge.
The dotted straight line shows the empirical red edge of the instability strip. Open circle show
stellar values taken from Table 1.}
\includegraphics{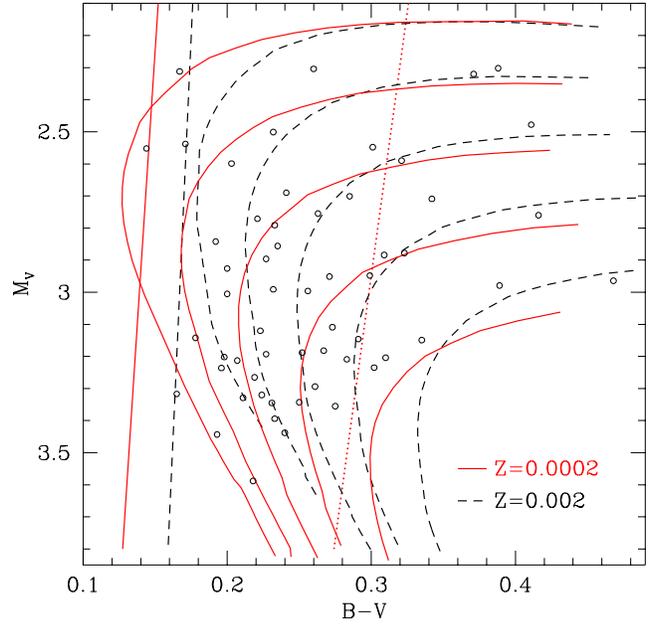}
\end{figure}

The blue edge was determined by means of our linear stability
analysis. In this plot, we chose the first overtone because this
mode is most frequently excited in the stars considered. The
fundamental mode blue edge is redder by some 0.022 mag in $B-V$
and that of the second overtone is bluer by 0.014 mag relative to
the that of the first overtone. These lines depend on mode
frequency but not on degree $\ell$. The adopted red edge is
empirical and based on $\delta$ Scuti data (Rodriguez et al.
2000).

We see in Fig. 4 that the blue edge is  sensitive to the assumed metal
content. Not surprisingly, the sensitivity to helium content is even
stronger. After all, it is helium that does most of the mode driving.
Increasing the helium abundance from $Y=0.25$ to 0.50 causes the
blueward shift of the first overtone blue edge by 0.05-0.06 mag.

Note that most of our stars fall within the instability strip.
However, there are few objects which are redder than the red edge
but admittedly their positions are quite uncertain. It may depend on
metallicity. There is only one star (NV321, not shown in the plot)
which is far bluer than the blue edge but, as we comment in the
next section, its colour is very uncertain. Thus, we see no
evidence for a substantial helium enhancement in stellar outer
layers.

The period-luminosity relations calculated for the first overtone
along the tracks are shown in Fig. 5. Similar relations for other
radial modes are obtained by the shifts given in the caption.
The relations depend on the colour and metal content. Thus,
application of SXPS as standard candles seems problematical.

We will use the plots shown in Figs. 4 and 5 to discuss the
parameters of individual objects and their mode identification.

\begin{figure}
\vspace{9cm}
\caption{The theoretical period -- luminosity relation for the
first overtone pulsation and the same models as in Fig. 4. The one for
the fundamental mode is shifted by +0.1 and that for the second overtone
by -0.08 in $\log P$.
Straight lines show corresponding positions of the first overtone blue edge.
}
\includegraphics{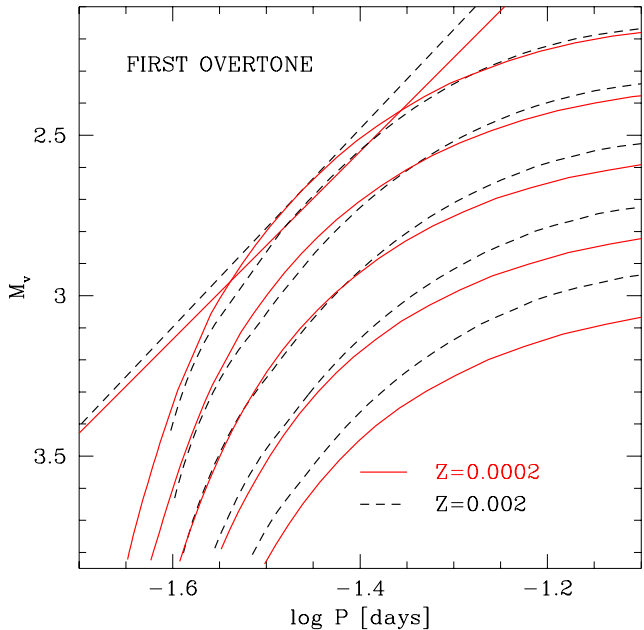}
\end{figure}

Fig. 6 shows the period ratios for consecutive radial mode pairs
as functions of the longer period in the pair (hereafter PRP
diagram). The relation for the first two radial modes is known as
the {\it Petersen Diagram}. We may see that, just as in the case
of double mode Cepheids, the $Z$-value is important but the
pattern of the period--period ratio relation is more complicated.

\begin{figure}
\vspace{10.4cm}
\caption{The theoretical period -- period ratio diagram for the first
four radial modes in our models. On the abscissa we plot the logarithm of the
longer period for each pair.}
\includegraphics{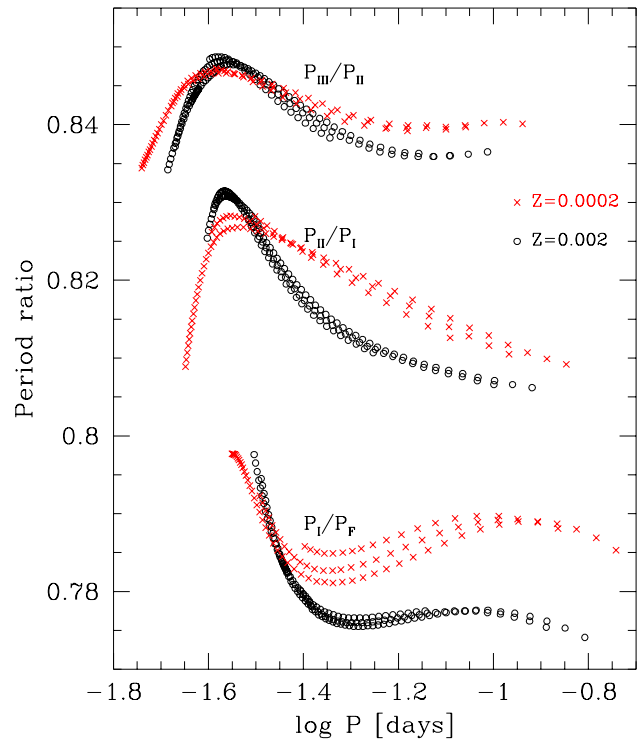}

\end{figure}

\section{Individual stars}

For each object, the first question we asked is whether any of the peaks
in the frequency spectrum could be identified with any of radial modes.
Stars with two radial modes exited are of special interest for
constraining stellar parameters. With this in mind we first asked
wheteher in our sample we find objects with the period ratios that are
consistent with the expected values for consecutive radial modes. Fig.
7, which is similar to Fig. 5 of Poretti (2003) for the high amplitude
Delta Scuti stars, shows that we have many candidates for two radial
mode pulsators. However, since the ranges for each period ratio are
rather wide, we cannot use Fig. 7 alone as a tool for a definite mode
identification. The witdh of the band for the specfic period ratio
arises from the uncertainty in metal abundance, in mass and in the
effective temperature. How these parameters affect the period shows is
shown in Fig. 6. Thus, both figures must be used jointly.

\begin{figure}
\vspace{10.4cm}
\caption{Selected period ratios for multiperiodic stars from Table 1,
which may fit the radial oscillations. Vertical lines indicate the
ranges of the theoretical period ratios for the first four radial modes
in our models, see Fig. 6.}
\includegraphics{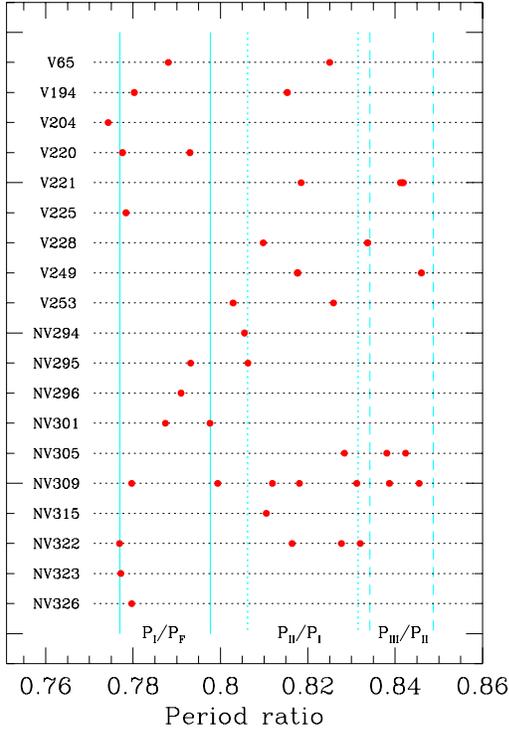}
\end{figure}

The agreement with the model values was regarded as a hint. If
none of the period pairs gave the ratio consistent with the values
seen in the PRP diagram, we concluded that at most only one peak
in the oscillation spectrum may be associated with a radial mode.
The case of high amplitude pulsation was considered as another
hint for the radial mode identification. It was then checked using
the position of the star in the H-R and PL diagrams. If there were
no high amplitude peaks, we checked only whether frequency of any
of the peak in the power spectrum agrees with any frequency of an
unstable radial mode (or its close nonradial neighbour) in the
model.

\begin{description}

\item {\bf V65}:~ The first SXPS found in the field of $\omega$ Cen. It
is out of our interest because it does not belong to the cluster
physically. The variable V65 is the star number 60026 in the proper
motion study of van Leeuwen et al. (2000). The probability of membership
for this variable is 0\%.

\item {\bf V194}:~ This is a very high amplitude pulsator (amp = 0.51
mag).  Positions of $f_0$ and $f_1$ modes on the PRP diagram (Fig.
6) show that these could be radial fundamental and first overtone
modes. According to the standard models the $Z$ value would be
intermediate between 0.002 and 0.0002. Then the mass would be
close to 1.05 $M_\odot$ according to its position in the H-R and
PL diagrams. Its evolutionary status would be near the end of core
hydrogen burning. The remaining peaks could be identified only
with nonradial modes. We emphasize that high frequency peak $f_2$
is still within the calculated instability range.

\item {\bf V195}:~ This is another high amplitude pulsator (amp = 0.38
mag). The shape of the light curve suggests the fundamental radial mode
identification.  Assuming standard evolutionary model, the implied mass
is close to 0.95 $M_\odot$ if $Z=0.0002$ and close to 1.10 $M_\odot$ if
$Z=0.002$. The two close peaks around the dominant frequency are
reminiscent of the pattern seen in Blazhko RR$ab$ stars.

\item {\bf V197}:~ There are three possible explanations of the
frequency pattern found in V197. First, the ratio between $f_0$ and
$f_3$ frequencies is 0.781 which may indicate that they correspond to
the radial fundamenal and first overtone modes. But the star is too red
for this hypothesis. The second possibility is that only one of three
highest peaks might correspond to the radial mode and it could be only
the second overtone and then the implied mass would be around 0.9
$M_\odot$ and $Z$ should be low. The two peaks close to the main
frequency may only be identified as nonradial modes. And alternatively,
the three modes might be interpreted as an $\ell=1$ triplet. Then the
rotational period would be around 12 days and rotational velocity would
be about 10 km/s.

\item {\bf V198}:~ The mean parameters are inconsistent with the value
of period. At this high temperature the period corresponds to a
high order $g$-mode which could not be excited. Perhaps it is a
low mass object. However, the color estimate is very uncertain in
this case (the error in $B-V$ is about 0.1 mag).

\item {\bf V199}:~ The highest amplitude SXPS in $\omega$ Cen. No other
significant frequencies were detected at the level above 0.005
mag. Undoubtedly, we see a fundamental mode pulsator. The implied
mass must be about 1.0 $M_\odot$ and the metallicity must be low.

\item {\bf V200}:~ The shape of the light curve points to the
fundamental radial mode identification. The second peak is
significantly lower, too low to consider the star as a classical
double-mode pulsator. The period ratio of 0.761 could be
consistent with the first overtone interpretation of $f_1$ only if
the metallicity is much higher than that considered for stars of
$\omega$ Centauri. Another problem is too high luminosity for the
period. Perhaps the object is a foreground Galactic star.

\item {\bf V204}:~ A high amplitude double-mode pulsator with period ratio
equal to 0.774 which indicates metallicity $Z=0.002$. Adopting
standard models, we get from the H-R and PL diagrams the mass of
1.15 $M_\odot$ and the shell hydrogen burning evolutionary status.

\item {\bf V217}:~ A monoperiodic and intermediate amplitude SXPS. In the
standard model with low metallicity and mass around 0.90 $M_\odot$ the
frequency corresponds to the first overtone of radial pulsation or
nearby nonradial mode.

\item {\bf V218}:~ The period ratio between two dominant modes falls
into the range of the 3rd to 2nd radial overtones ratio. However,
for standard models, this identification is inconsistent with the
position in the H-R and PL diagrams, which indicates that $f_0$
could be only the fundamental or a nearby nonradial mode. Such an
identification implies mass range $0.95 \div 1.00$ $M_\odot$ and
$Z$ around 0.0002. Another possibility is that the star is
significantly undermassive.

\item {\bf V219}:~ The power spectrum of the star is very much like that
of a typical main sequence $\delta$ Scuti star, that is, there are many
low amplitude peaks spread over wide range of frequencies. Such spectra
are extremely difficult to interpret. The main frequency is consistent
with fundamental radial mode identification if the metallicity is low
and the mass is around 1.00 $M_\odot$. The star position in the H-R
diagram is near ZAMS, which indicates that the star could be a recent
merger of two stars of comparable mass. The low frequency $f_3=3.28$ c/d
is troublesome. Definitely too low for $p$-modes. The star is unlikely to
excite $g$-modes, such as seen in $\gamma$ Doradus stars, because it is
one of the hottest in our sample. Perhaps we see ellipsoidal light
variations. This is an interesting possibility with ramifications to the
problem of BSS origin. The implied orbital period of 0.609 d would be
significantly shorter than any of the BSS in the Preston and Sneden
(2000) list.

\item {\bf V220}:~ A moderate amplitude multimode pulsator with period
ratio between two dominant frequencies equal to 0.778. It implies
metallicity close to $Z=0.002$, which is also consistent with the
position of the star in the H-R and PL diagrams and leads to mass
estimate of 1.10 $M_\odot$. There is a close neighbour of the {\it bona
fide} 1st overtone and it may be interpreted only in terms of nonradial
modes.

\item {\bf V221}:~ Many low amplitude peaks, none two of them may be
interpreted as two radial modes. High frequencies combined with $M_V$
value indicate excitation of high order $p$-modes.

\item {\bf V225}:~ A high amplitude pulsator (amp = 0.22 mag) with period
ratio between two dominant frequencies equal to 0.778. It implies
metallicity close to $Z=0.002$, which is also consistent with the
position of the star at the H-R and PL diagrams and leads to mass
estimate of 1.15 $M_\odot$. There is a close neighbour of the 1st
overtone candidate which could be interpreted as nonradial mode.

\item {\bf V226}:~ A high amplitude pulsator (amp = 0.17 mag). Frequency
$f_0=26.4185$ c/d may correspond to the first overtone, if the
metallicity is rather high and the mass is around 1.05 $M_\odot$.
Then the frequencies $f_1$ and $f_3$ close to $f_0$ could be
associated with nonradial modes. The frequency $f_2$ is too low to
correspond to a nonradial mode located in the neighborhood of the
radial fundamental mode; therefore, it could be one of the pure
$g$-modes.

\item {\bf V227}:~ Frequency $f_0=26.1605$ c/d may correspond to the
first overtone according to the positions of the star in the H-R
and PL diagrams. The implied mass estimate is 1.1 $M_\odot$ and
the metallicity  around $Z=0.002$.  Frequencies $f_1$ and $f_2$
could be nonradial modes in the vicinity of the first overtone
radial mode.

\item {\bf V228}:~ A moderate amplitude multimode pulsator with period
ratio between the frequencies $f_0$ and $f_3$ equal to 0.834.
Looking at the PRP diagram one can find that they can be the first
and second overtone radial modes but only in the case of low
metallicity. The frequencies $f_1$ and $f_2$ correspond then to
nonradial modes in the vicinity of the first overtone. Position of
the star on the H-R and PL diagrams agrees with assumption of a
low metallicity and indicates the mass of 0.90 $M_\odot$.

\item {\bf V229}:~ It is one of the bluest SXPS in our sample, much
bluer than Kaluzny et al. (1996) found. Since they relay on $V-I$
colours, we checked the object carefully for blending effects and found
none. Our colour implies that the dominant period is too long to
correspond to the fundamental mode. The problem which is similar to
the cases of V198 and V226, where we do not expect excitation of pure
$g$-modes in so hot star.

\item {\bf V231}:~ The dominant frequency $f_0=26.6778$ c/d could be due
to the  radial fundamental mode only for $Z=0.0002$. In this case, the
mass of the star is between 0.95 and 1.00 $M_\odot$. The frequencies
$f_1$ and $f_2$ could be due the excitation of two nearby nonradial
modes. The frequency $f_3=36.6960$ c/d is associated with nonradial mode
placed between the first and the second overtone.

\item {\bf V233}:~ The dominant frequency $f_0=27.3691$ c/d could be due
to the radial first overtone mode only if $Z\approx 0.0002$. In
this case, the mass of the star would be around 0.95 $M_\odot$. The
low frequency peak at $f_1=2.9205$ c/d could indicate that the
star is, along with V219, a candidate for an ellipsoidal variable.

\item {\bf V237}:~ The high amplitude (0.27 mag) peak at $f_0=15.2433$
c/d could be identified as the radial fundamental mode. Such an
identification is consistent with the positions in the H-R and PL
diagrams for low $Z$ models. Then the mass of the star is about 0.95
$M_\odot$. If $f_0$ is indeed the fundamental mode then the expected
position of the second overtone radial mode is around 24 c/d. Thus, we
conclude that the frequency $f_1=24.1162$ c/d could correspond to the
second overtone. The peak at $f_2=15.2013$ c/d is due to a nonradial
mode in the vicinity of fundamental frequency.

\item {\bf V238}:~ Again, the dominant frequency $f_0=24.5096$ c/d could
be due to the excitation of the radial fundamental mode only if $Z$ is
low and the mass of the star is 0.95 $M_\odot$. The frequencies $f_1$,
$f_2$ and $f_3$ may be attributed to nearby nonradial modes. The
expected position of the second overtone radial mode is around 38 c/d
thus the frequency $f_4=39.5205$ c/d would be due to the nonradial
$p$-mode.

\item {\bf V249}:~ A moderate amplitude multimode pulsator with the period
ratio between two dominant frequencies equal to 0.818. It is
consistent with an assumption that $f_0$ and $f_1$ correspond to
the first and second overtone radial modes. This requires high $Z$
values, somewhat higher than $Z=0.002$. However this
interpretation is inconsistent with the star position in the H-R
and PL diagrams. 

\item {\bf V250}:~ A double mode pulsator with $f_0$ and $f_1$ being the
radial fundamental and the second overtone modes, respectively.
The position at the H-R and PL diagrams indicates a low
metallicity and the mass between 0.90 and 0.95 $M_\odot$.

\item {\bf V252}:~ A monoperiodic and low amplitude star. The position
of the star in the H-R and PL diagrams is consistent with the
fundamental radial mode identification only for a low metallicity
and the mass between 0.85 and 0.90 $M_\odot$.

\item {\bf V253}:~ A moderate amplitude multimode pulsator. The dominant
frequency could be the fundamental radial mode only for our low
metallicity models. In this case, the star has mass around 1.00
$M_\odot$. The frequency $f_1$ corresponds to a nonradial mode in
the vicinity of fundamental peak and frequencies $f_2$ and $f_3$
to some high order nonradial $p$ modes.

\item {\bf NV294}:~ This is the shortest period SXPS known.  Up to now,
the shortest period variable of this type was V10 from NGC6397 (Kaluzny
and Thompson 2003). NV294 is a multimode and very low amplitude object.
With our H-R and PL diagrams the frequency $f_1=28.9864$ c/d may be
associated with the radial fundamental mode, in a model of high
metallicity and mass around 1.2 $M_\odot$. In this picture, the
frequencies $f_0$, $f_2$ and $f_3$ correspond to the high order nonradial
$p$-modes.

\item {\bf NV295}:~ This is another very short period, low amplitude and
multimode object. The frequency $f_1=28.3558$ c/d could correspond to
the radial second overtone mode but only for low values of $Z$. In this
case, the position of the star at the H-R and PL diagrams indicates the
mass between 0.90 and 0.95 $M_\odot$. The frequencies $f_0$, $f_2$ and
$f_3$ could correspond to nonradial modes.

\item {\bf NV296}:~ Another object similar to NV294 and NV295. There are
five frequencies present in the light curve of the star. With our models
only the lowest amplitude peak at frequency $f_4$ might be identified
with a radial mode.  It would be the second overtone and in this case
the metallicity of the star would be around $Z=0.002$ and the mass would
be between 1.10 and 1.15 $M_\odot$. In such a case, other frequencies
would correspond to nonradial high order $p$-modes.

\item {\bf NV297}:~ A low amplitude multimode object with four relatively
close peaks. One of them could be identified with the second overtone
radial mode and then the star mass would be around 1.20 $M_\odot$ at
$Z=0.002$ and about 1.05 $M_\odot$ at $Z=0.0002$. We consider
possibility that three of the four frequencies represent a rotationally
splitted $\ell=1$ triplet. In the plausible interpretation $f_1$ would
be a retrograde and $f_3$ would be the prograde mode. Then the frequency
separation would imply rather fast rotation (over 100 km/s). However,
the calculated position of the $m=0$ component does not agree with the
position of any of the two remaining peaks. It was much closer to the
$f_3$.

\item {\bf NV298}:~ The dominant peak of the two observed in the star
may be identified, in our models, with the fundamental radial mode
if the metallicity is rather high and the mass is between 1.15 and
1.20 $M_\odot$. The second close frequency must then be associated
with a nonradial mode.

\item {\bf NV299}:~ The dominant peak could be the radial first
overtone  mode in a star of a relatively high metallicity and mass
around 1.05 $M_\odot$. Then the frequency $f_1$ would correspond to a
nonradial mode located between the radial first and second overtones.
The star is slightly too red for this interpretation but it is located
in a crowded region of the cluster and its $V$ magnitude zero point is
determined with the accuracy of around 0.04 mag. Thus the star may be in
fact even 0.05 mag bluer.

\item {\bf NV300}:~ A monoperiodic and low amplitude SXPS. The star could
be the radial fundamental mode pulsator only if we assume the high
metallicity and the mass of around 1.15 $M_\odot$.

\item {\bf NV301}:~ A multimode and low amplitude SXPS. One of the two
dominant frequencies may be the radial second overtone mode,
according to our models. The star has then a high metallicity and
the mass around 1.05 $M_\odot$. The frequencies $f_2$ and $f_3$
should correspond to the high order nonradial $p$-modes.

\item {\bf NV302}:~ A monoperiodic and low amplitude star. The frequency
$f_0=28.1538$ c/d could be the first overtone radial mode but only for
our $Z=0.002$ models. In this case, the mass of the star is between 1.10
and 1.15 $M_\odot$. It is also possible that $f_0$ is the second
overtone. This requires low $Z$ and, in this case, the mass of the star
is between 0.90 and 0.95 $M_\odot$.

\item {\bf NV303}:~ A monoperiodic and very low amplitude star. The only
frequency determined could be associated with the first radial overtone.
Then the star has low metallicity and mass slightly less than 1.0
$M_\odot$.

\item {\bf NV304}:~ A similar pulsator to NV303. Again, the first radial
overtone interpretation possible, however in this case the models imply
high metallicity and mass around 1.05 $M_\odot$.

\item {\bf NV305}:~ A multiperiodic object with interesting power
spectrum. The amplitude increases systematically with the frequency. The
period ratio between two dominant peaks is equal to 0.828 - a value
which could be the ratio between the first and second radial overtones,
but only for very low metallicity. It is supported by the position of
the star in the H-R and PL diagrams implying the mass in range of
$0.85\div 0.90~ M_\odot$. The frequencies $f_2$ and $f_3$ would
correspond to nonradial peaks in the vicinity of the fundamental radial
mode.

\item {\bf NV306}:~ One of the faintest SXPS in our sample. The dominant
frequency could be due to the radial fundamental mode. With this
interpretation, the metallicity of the object is low and its mass is
between 0.90 and 0.95 $M_\odot$. The expected position of the second
overtone radial mode is close to 40 c/d, thus the second peak could be
connected with this mode.

\item {\bf NV307}:~ A monoperiodic and moderate amplitude star. Our models
imply that the excited mode must be of higher order than two, which
seems to be in conflict with the star relatively low temperature.

\item {\bf NV308}:~ A multiperiodic object. Of the three dominant peaks at
low frequency one could be identified with the first overtone for
our low metallicity models with the mass of about 0.90 $M_\odot$.
We also considered the possible interpretation of these three
peaks in terms of rotationally splitted $\ell=1$ mode. In this
case, the problem is opposite to that accounted in NV297 - the
observed asymmetry is too high for the rotation rate of about 50
km/s implied by the distance between $f_0$ and $f_1$.

\item {\bf NV309}:~ The star has one of the richest frequency spectrum.
We determined as many as seven frequencies. The $f_5/f_1$ ratio could be
the frequency ratio between the first and second overtones of radial
pulsations. This interpretation is however in conflict with the H-R
position, where the star is too red. For this star we obtained a
consistent interpretation of three peaks ($f_5$, $f_0$ and $f_3$) as a
rotational splitted $\ell=1$ triplet. For our model with $Z=0.0002$ and
mass of 0.95 $M_\odot$ the separation between $f_3$ and $f_5$ implies
rotational velocity of 60 km/s. With this velocity the centroid
component, displaced from the center by the second order effect of
rotation, comes close to $f_0$. This triplet is not far from the
frequency of the first radial overtone. The peak $f_4$ would then be
located in the vicinity of the second overtone while the highest
frequency peaks ($f_2$ and $f_1$) of the third overtone.

\item {\bf NV310}:~ The only two frequencies are close to the expected
position of the radial second overtone, for our models with high
metallicity and mass around 1.10 $M_\odot$.

\item {\bf NV311}:~ The two frequencies measured in this star,
$f_0=24.1469$ c/d and $f_1=36.6940$ c/d, may be due to the radial
fundamental and the second overtone modes, respectively. Unfortunately,
we can not verify this hypothesis because we do not know the exact
magnitude and color of the star, which is located in very crowded
region of the cluster.

\item {\bf NV312}:~ One of the brightest SXPS in $\omega$ Cen. The
frequency $f_0=23.0802$ c/d could be associated with the radial second
overtone mode but only for high metallicity. In this case, the mass of
the star is around 1.2 $M_\odot$. The peak at $f_1=36.8$ c/d can be
interpreted as the high order $p$-mode.

\begin{figure*}
\vspace{9cm}
\caption{Phased light curves of monoperiodic SXPS in $\omega$ Centauri}
\includegraphics{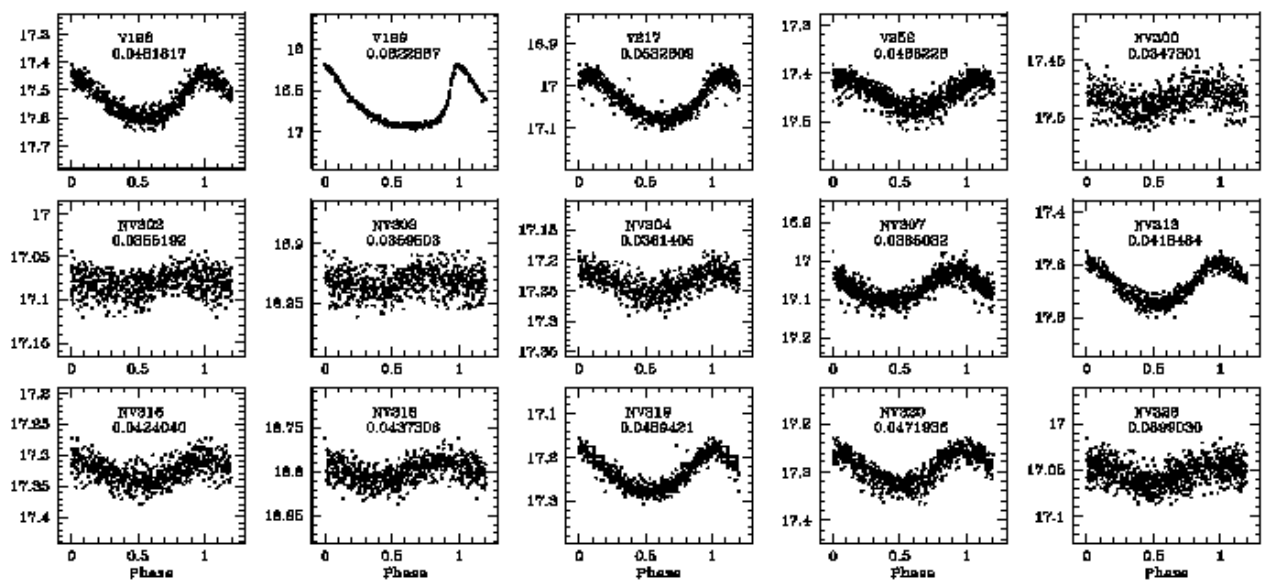}
\end{figure*}

\item {\bf NV313}:~ A monoperiodic, high amplitude and faintest SXPS in
the field of $\omega$ Cen. The high amplitude may suggest that the
star is a radial mode pulsator. The problem is that in our
standard models, the star with the observed effective temperature
and luminosity would have the fundamental radial mode period by
factor 1.3 shorter. The star is also one of the hottest in our
sample and we certainly do not expect excitation of $g$-modes in
such a star. In fact, we would rather expect higher order modes
excitation, therefore this object is likely to be undermassive by
factor of 1.7 or more.

\item {\bf NV314}:~ The frequencies $f_0=23.7405$ c/d and $f_2=36.4036$
c/d could be the radial fundamental and second overtone modes,
respectively, but only if the star would be bluer by 0.06 mag. It is
possible, because the star is located in the crowded region (three
bright companions) and zero point in $V$-band is poorly determined.

\item {\bf NV315}:~ One of the brightest SXPS in $\omega$ Cen with a
rich oscillation spectrum. The photometric data combined with our model
parameters imply that even the lowest frequency, $f_0$, is higher than
the second overtone radial mode. The preference to the high frequency
mode excitation seems to be in conflict with the low temperature of the
star, the one of the lowest in our sample.

\item {\bf NV316}:~ A monoperiodic and low amplitude star. The period is
too long for fundamental radial mode, according to our models. Perhaps
the star is another case of an undermassive object.

\item {\bf NV317}:~  A low amplitude multiperiodic star. Assuming our
standard models, the frequency $f_0=23.4525$ c/d could be attributed to
the radial second overtone mode in a star with a rather high metallicity
and mass of around 1.05 $M_\odot$. Three other peaks ($f_1$, $f_3$ and $f_2$)
in this model can be interpreted as a rotationally splitted $\ell=1$ triplet,
which implies rotational velocity of about 120 km/s.
Alternatively, another combination of three peaks
($f_0$, $f_3$ and $f_2$) might be interpreted as an $\ell=1$
triplet, which would imply rotational velocity of about 80 km/s.

\item {\bf NV318}:~ A low amplitude and monoperiodic star. The frequency
$f_0$ could be due to the radial second overtone mode in a model with
high metallicity and mass between 1.05 and 1.10 $M_\odot$.

\item {\bf NV319}:~ A moderate amplitude and monoperiodic star. The only
frequency could be that of the radial first overtone mode in a star of
low metallicity and the mass between 0.85 and 0.90 $M_\odot$.

\item {\bf NV320}:~ It is almost identical to the previous object.

\item {\bf NV321}:~ A multiperiodic and moderate amplitude star. The
frequencies $f_0$, $f_1$ and $f_2$ form an asymmetric triplet which
perhaps might be interpreted as rotationally splitted $\ell=1$ mode.
Unfortunately, the variable is located in very crowded field and the
mean $B$ and $V$ magnitudes are very uncertain.

\item {\bf NV322}:~ A moderate amplitude variable with one of the richest
power spectrum. The ratio between two dominant frequencies is 0.777
which may suggest that they are fundamental and first overtone radial
modes. We have no color determination for this star and we can not
exploit this hypothesis.

\item {\bf NV323}:~ A double mode and low amplitude pulsator.
Surprisingly, we found the same period ratio of 0.777 as in previous
case. However, we have a problem with interpretation of the frequencies
in terms of the fundamental and first overtone radial pulsations. The
period ratio suggests the metallicity of $Z\approx 0.002$. Our models
imply the H-R position by some of 0.06 mag bluer than observed.

\item {\bf NV324}:~ A high amplitude and multiperiodic star. Its rich
power spectrum is shown in Fig. 2. According to our models, the dominant
frequency is close to the radial first overtone for low metallicity and
mass of around 1.05 $M_\odot$. The frequencies $f_1$, $f_2$ and $f_5$
are all close to that of the third radial overtone mode.

\item {\bf NV325}:~ A moderate amplitude and multiperiodic object. The
dominant frequency may be interpreted, according to our models, as due
to the second radial overtone in a star with  mass between 1.15 and 1.20
$M_\odot$ if $Z=0.002$ or between 1.00 and 1.05 $M_\odot$ if $Z=0.0002$.
Presence of three high frequency modes is surprising in view of rather
low temperature and high luminosity of the object.

\item {\bf NV326}:~  This is a classical double mode star with
$f_0=17.5728$ c/d as radial fundamental and $f_1=22.5375$ c/d as the
first overtone modes. The frequency ratio of 0.780 nicely agrees with
that of fundamental and first overtone in an intermediate metallicity
object. Further, the periods and the metallicity agree with the position
of the star in the H-R and PL diagrams. The implied mass is then between
0.95 and 1.00 $M_\odot$.

\item {\bf NV327}:~ It is one of the bluest stars in our sample. The
frequency $f_0$ may be associated with the fundamental mode but only
if the star is significantly undermassive.

\item {\bf NV328}:~ A low amplitude and monoperiodic star. It is the
reddest star in our sample. The only frequency, according to our models,
may be interpreted as the fundamental radial mode, implying the mass
around 1.00 $M_\odot$ at $Z=0.0002$ or mass between 0.85 and 0.90
$M_\odot$ at $Z=0.002$.

\end{description}

\section{Conclusions}

In most cases the observed properties of SXPS in $\omega$ Centauri
may be explained in terms of standard evolutionary models and
their linear nonadiabatic properties. In a number of cases, where
the observational evidences pointed to radial mode excitation, we
derived the masses of the objects. They all fall into the range
$0.90 \div 1.15 M_\odot$. The implied evolutionary status covers
both the main sequence and early post-main sequence evolutionary
phases.

Only in four cases the observed parameters were inconsistent with the
standard models and could be reconciled with the undermassive objects.
Such objects could arise in several situations. They can appear as a
result of mergers, if there is a mass loss or element mixing. They could
also arise as the result of mass exchange in binary systems. This
possibility may be contemplated only in the case of very large mass loss
on the lower red giant branch. Applicable models would have rather low
mass ($\sim 0.2 M_\odot$), most of which contained  in the helium core.
Such models were once considered by Dziembowski and Koz{\l}owski (1974)
for high amplitude $\delta$ Scuti stars, called then Dwarf Cepheids.

We found two cases of stars with long periodicities that may be
interpreted as due to the tidally induced distortion. This could
be an indirect evidence for binarity. Possible implications for
the mechanism of BSS formation are interesting. Therefore, the two
objects deserve further studies to check the interpretation. This
could be done by searching for induced pulsation period changes
(current data do not allow to reject nor confirm the proposal) or
by means of radial velocity measurements.

Our search for rotationally split triplets resulted in three plausible
cases. For NV309 the inferred equatorial velocity would be
about 60 km/s, and for NV317 it would be about 80 or 120 km/s,
depending on three frequency peaks involved.
For V197 the interpretation of all three close peaks as $\ell=1$ triplet
results in equatorial rotational velocity of about 10 km/s.

We found considerable diversity in pulsational behaviour ranging from
that typical for main sequence $\delta$ Scuti stars characterized by low
amplitude and multimode variability to high amplitude Cepheid-like
monomode pulsations. This is quite amazing bearing in mind the fact that
the spread in $V$ brightness is only about one magnitude. To give the
picture of the observed diversity, Fig. 8 shows the phased light curves
of the monoperiodic pulsators from our sample.

Examples of $\delta$ Scuti-like behaviour are objects like V219,
NV294, NV295 and NV296. These are among the faintest and the
hottest stars in our sample. Amongst the faintest objects there
are also some relatively high amplitude pulsators (V198, V249 and
NV313) but all of them are suspected for being undermassive.

The star V199, whose light curve resembles those of Cepheids or
RR$ab$ stars, is amongst most luminous objects. Other high
amplitude objects are multiperiodic. Among them there are five
classical double mode pulsators i.e. stars with fundamental and
the first overtone excited. These objects were crucial for
deriving constrains on masses and metallicities. In three of them
we detected additional, apparently nonradial modes. Coexistence of
high amplitude radial modes with low amplitude nonradial modes was
first seen by Walraven, Walraven and Balona (1992) in the $\delta$
Scuti star AI Vel (see also Poretti 2003). In some of our cases
the nonradial modes are located close to radial modes in the
frequency spectrum but not in all, just like in AI Vel.

Three clean cases of spectra consisting of a dominant peak
surrounded by close low amplitude peaks are V195, V197 and NV321.
These are possible analogs of Blazhko RR$ab$ stars.

\section*{Acknowledgments}

We are thankful to Dr. Poretti, the referee, for pointing
mis-identification of $f_0+f_1$ frequency in NV324. This work was
supported by the KBN grants number 1~P03D~006~27 to A. Olech,
5~P03D~030~20 to W.A. Dziembowski,  5~P03D~012~20 to A.A.
Pamyatnykh, and 5~P03D~004~21 to J. Kaluzny. A.
Schwarzenberg-Czerny would like to acknowledge generous grant by
Clifford and Mary Corbridge Trust of Cambridge, England.

\end{document}